# Edge-guided inverse design of digital metamaterial-based mode multiplexers for high-capacity multi-dimensional optical interconnect

Aolong Sun[1,2,3,†], Sizhe Xing[1,2,†], Xuyu Deng[1,2,3], Ruoyu Shen[1,3], An Yan[1,2], Fangchen Hu[3], Yuqin Yuan[1,2], Boyu Dong[1,2], Junhao Zhao[1,2], Ouhan Huang[1,2], Ziwei Li[1,2], Jianyang Shi[1,2], Yingjun Zhou[1,2], Chao Shen[1,2], Yiheng Zhao[3], Bingzhou Hong[3], Wei Chu[3,*], Junwen Zhang[1,2,*] Haiwen Cai[3] &Nan Chi[1,2,*]

[1]School of Information Science and Technology, Fudan University, Shanghai, China

[2]Key Laboratory for Information Science of Electromagnetic Waves (MoE), Fudan University, Shanghai, China

[3]Zhangjiang Laboratory, Shanghai, China

[†]*These authors contributed equally to this work.*

[*]*Corresponding authors:* junwenzhang@fudan.edu.cn, chuwei@zjlab.ac.cn ,nanchi@fudan.edu.cn

## Abstract

The escalating demands of compute-intensive applications urgently necessitate the adoption of optical interconnect technologies to overcome bottlenecks in scaling computing systems. This requires fully exploiting the inherent parallelism of light across scalable dimensions for data loading. Here we experimentally demonstrate a synergy of wavelength- and mode- multiplexing combined with high-order modulation formats to achieve multi-tens-of-terabits-per-second optical interconnects using foundry-compatible silicon photonic circuits. Implementing an edge-guided analog-and-digital optimization method that integrates high efficiency with fabrication robustness, we achieve the inverse design of mode multiplexers based on digital metamaterial waveguides. Furthermore, we employ a packaged five-mode multiplexing chip, achieving a single-wavelength interconnect capacity of 1.62 Tbit $s^{-1}$ and a record-setting multi-dimensional interconnect capacity of 38.2 Tbit $s^{-1}$ across 5 modes and 88 wavelength channels, with high-order formats up to 8-ary pulse-amplitude-modulation (PAM). This study highlights the transformative potential of optical interconnect technologies to surmount the constraints of electronic links, thus setting the stage for next-generation datacenter and optical compute interconnects.

## Introduction

The exponential growth of hyper-scale datacenters[1–3], driven by the intensifying computational demands of applications such as artificial general intelligence[4], cloud computing[5], and virtual reality[6], urgently necessitates a shift from traditional copper wire-based links. On-chip optical interconnect technology[1,2,7] is currently establishing itself as pivotal for ultra-high-capacity data exchange between multi-node, distributed computing systems[8], offering significant advantages in power efficiency and latency reduction over electrical counterparts. Crucially, the integration of mode division multiplexing (MDM), dense wavelength division multiplexing (DWDM), and advanced modulation formats is essential to fully exploit the multi-dimensional capabilities of light for data loading[9–12]. This triple integration allows massively parallel transmission of multiple orthogonal modes within a single waveguide, each capable of supporting an extensive array of wavelength channels with high spectral efficiency, as depicted in Fig. 1. This multimode waveguide-based configuration substantially reduces the shoreline density of on-chip optical interconnects compared to conventional single-mode waveguide bundles, and can further decrease the required cross-sectional area when integrated with multimode fibers.

As high-performance integrated optical frequency combs[13–16] and multi-wavelength transmitters based on tunable microring modulators[17–19] evolve, there is a critical need for innovative chip-based, ultra-compact high-order mode multiplexers (MUX) to achieve broad operation bandwidth, micrometer-scale dimensions, and low-crosstalk mode multiplexing capabilities. Although phase-matching principles have facilitated broadband and high-order MUXs capable of efficiently converting optical signals into multiple orthogonal modes[20–25], these designs require cascading several mode converters, substantially increasing their footprint with higher mode counts. Additionally, phase-matching MUXs are prone to horizontal fabrication errors, complicating large-scale production and dense integration[23]. To address these challenges, it is necessary to move beyond intuition-first design approaches and analytical theories. Inverse design, as an emerging paradigm in goal-oriented design[26–32], enables the automatic iterative optimization of a large number of design parameters within predefined numerical boundaries through advanced mathematical algorithms. Furthermore, all-dielectric metamaterials[33,34], characterized by their extraordinary ability to manipulate electromagnetic waves within numerous sub-wavelength structures, are particularly amenable to inverse design methods. Inverse-designed all-dielectric metamaterials have been proven effective in a range of integrated photonic applications, such as optical communication[20,26,28,32,35], computing[27], sensing[29], and imaging[31]. Inverse-designed metamaterials can be categorized into digital and analog types based on their permittivity distribution patterns[36]. Analog metamaterials (AMs) typically display irregular patterns with curvilinear boundaries and isolated islands, while digital metamaterials (DMs) feature regularized etching patterns, such as squares or circles. The inverse design of complex devices, such as high-order mode MUXs, is impractical with the brute-force search algorithms commonly employed for DMs[9,32,35,37–40]. In contrast, the gradient-based methods for AMs offer enhanced optimization efficiency and expanded design freedom. However, the challenges posed by the small minimum feature size (MFS) and irregular shapes of AMs present significant fabrication hurdles. Additionally, the reactive ion etching (RIE) lag effect causing variable etching depths for different-sized holes[41], further complicates the accurate replication of the designed local effective index. To date, robust AM designs have only been demonstrated in simple standalone devices with few ports[42–44], while DM designs have found broader applications in integrated photonic circuits[9,35,39,45]. Due to the outlined challenges, inverse-designed mode MUXs fabricated on the silicon-on-insulator (SOI) platform have been reported to support a maximum of only four modes[26,35,38,45–47].

Therefore, an advanced inverse design method that can merge the advantages of both analog and digital metamaterials is vital for fabricating complicated and robust photonic devices. In this paper, we introduce an edge-guided analog-and-digital optimization (EG-ADO) method, which combines the computational efficiency of topology optimization (TO) with the fabrication reliability of DMs. This is achieved by innovatively integrating edge detection algorithms into the conversion process from analog to digital metamaterials. This method enables the design and experimental demonstration of broadband multimode multiplexers (MUXs) on SOI, including a four-mode MUX ($TE_0$–$TE_3$), a five-mode MUX ($TE_0$–$TE_4$), and a six-mode MUX ($TE_0$–$TE_5$) for scalability validation. Building upon the packaged five-mode MDM chip characterized by exceptional uniformity, low loss, and minimal crosstalk, we achieve a MIMO-free high-speed single-$\lambda$ on-chip interconnect with a line rate exceeding 1.6 Tb $s^{-1}$ and a record per wavelength per mode channel rate of 324 Gb $s^{-1}$. By combining DWDM with MDM techniques, we further realize a multi-dimensional ultra-high-capacity on-chip interconnect, supporting 440 channels with a record aggregate net rate exceeding 38 Tb $s^{-1}$ and a high spectral efficiency of 8.68 bits $s^{-1}Hz^{-1}$. Finally, the compatibility of the proposed digital metamaterials with mainstream foundry processes highlights the potential for large-scale and cost-effective deployment in future optical compute interconnect applications (Fig. 1), potentially supporting the 32T and even the 64T generations[48].

## Results

### Principle and performance of the EG-ADO method

We illustrate the principle and performance of the proposed EG-ADO method through its application in designing a fabrication-robust DM-based five-mode MUX. The initial schematic of this MUX is depicted in Fig. 2(a), consisting of a metamaterial design region spanning 10×6 μm². Upon optimization, the MUX can effectively transform the fundamental TE modes ($TE_0$) from the five single-mode waveguides into the first five TE modes ($TE_0$-$TE_4$) in the multimode waveguide. Due to reciprocity, this MUX also serves effectively as a mode demultiplexer (deMUX) by swapping the input and output ports. The material platform is the standard 220 nm SOI with $SiO_2$ cladding and buried oxide (see Methods for detail). The finite-difference eigenmode (FDE) solver from ANSYS Lumerical is utilized to calculate the effective index of various orthogonal modes as a function of waveguide widths (see Fig. S4). The widths of the single-mode waveguide and the multi-mode waveguide are set at 500 nm and 3000 nm, respectively.

The EG-ADO method consists of three main stages, as illustrated in Fig. 1(a). The first stage involves iterative topology optimization of the permittivity distribution within the metamaterial design region using the adjoint method. To accurately calculate the pixel gradients, the pixel size of the design area is set to 20×20 nm$^2$. During each iteration, we obtain the electric field distributions of the metamaterial design region through one forward and one adjoint electromagnetic simulation for each mode. These fields are used to compute the gradient of the objective function (see Methods) with regard to the permittivity distribution. The permittivity value for each pixel is then updated continuously between $\varepsilon_{Si}$ and $\varepsilon_{SiO_2}$ using gradient-based optimizer to enhance the overall device performance. After multiple iterations, an AM-based mode MUX demonstrating optimal performance is obtained. However, challenges in reliable fabrication arise due to the incomplete binary conversion of the index pattern and the presence of notably small gaps and holes, which complicate manufacturing processes. To address these fabrication challenges, the second stage performs an edge-guided conversion from analog to digital metamaterials. This is achieved by extracting critical edge information from the material boundaries within the analog pattern using the Canny edge detection algorithm[49]. To accommodate the MFS constraints of commercial foundries, the extracted edge map undergoes dimensional reduction through a max-pooling operation (see Supplementary Note 1), enlarging the pixel size from 20 nm to 120 nm. Following this, the feature map containing edge information is binarized to produce a decision map that guides the subsequent index mapping process. During index mapping, non-edge pixels (black) in the decision map directly adopt the index values from their corresponding positions in the original analog pattern, while the state of edge pixels (white) is temporarily designated as to-be-determined (TBD). Notably, TBD pixels constitute only a small fraction of all pixels, enabling the design to be finalized with a minimal number of iterations in the next stage, thus significantly conserving computational resources. In the final stage, a customized direct binary search (DBS) algorithm identifies the material for each TBD pixel in the converted pattern. For these TBD pixels, the refractive index is alternately set to that of Si or $SiO_2$, and the objective function is computed for each setting. The material that results in superior performance is then confirmed for each pixel. This stage terminates after a single pass through all TBD pixels, finalizing the design of the DM-based five-mode MUX. The inset (ii) in Fig. 1(a) distinctly shows that throughout the digital optimization process, irregular gaps and holes within the metamaterial design region are progressively replaced by uniform squares in a column-wise direction. The evolution of device performance during the entire optimization process is provided in Supplementary Note 1.

The final structure of the DM-based five-mode MUX is given in Fig. 2(b), featuring a large MFS which is well-suited for ultraviolet lithography (UVL) fabrication. Insertion loss is quantitatively analyzed by evaluating the conversion efficiency from the incident mode to the target mode based on mode overlap integrals (Fig. 2(c)). Inter-mode crosstalk is assessed by computing the mode overlap between the obtained mode profiles and other undesired orthogonal modes. The reflection and scattering are also discussed in the Supplementary Note 3. The numerical results demonstrate that this device maintains a significantly flat response (see Fig. S5), with loss below 1.96 dB and crosstalk under −15.81 dB across the entire C band. Fig. 2(d) demonstrates the manipulation of optical field by finely etched holes within the metamaterial design area.

To further explore the generalizability and efficiency of the proposed EG-ADO method and its impact on devices of varying complexities, we have applied this method to design two-, four-, and six-TE-mode MUXs (details in Supplementary Notes 3 and 7). Fig. 2(e) compares the computational complexity of various inverse-designed DM-based mode MUXs[9,35,40,45], measured by the required number of Maxwell solves (RNMS). The fitting results show that the RNMS for the DBS method scales quadratically with the number of modes, leading to excessive resource demand for high-order designs. In contrast, devices designed using the EG-ADO method show a linear scaling of RNMS (see Methods), with more pronounced improvements as the mode number increases. Additionally, for the proposed EG-ADO method, it is noteworthy that increasing the number of TBD pixels by thickening the edges (see Supplementary Notes 1 and 3) does not correlate with an evident performance gain, as shown in Fig. 2(f). Here, the conversion rate is defined as the proportion of pixels directly undergoing index mapping. This observation highlights the critical role of edge pixels and indicates a diminished return on optimizing pixels distant from edges, thus reducing unnecessary computational costs.

## Optical performance of fabricated devices

We fabricate the four-, five-, and six-mode MUXs on the SOI platform using electron beam lithography (EBL), with detailed processes provided in Methods. Due to the absence of imaging equipment, we assess the optical performance by cascading two identical devices back-to-back on the chip. The fabricated back-to-back four-mode MDM system (Fig. 3(a)) includes a pair of four-mode MUX and deMUX linked by a multimode waveguide. The multimode waveguide in the four-mode MUX has a width of 1.8 μm, and the design region occupies an area of 6×6 μm². The coupling loss of the grating couplers is approximately 5.7 dB/facet at the peak wavelength. The transmission spectra of the four-mode MDM circuit are measured and normalized by subtracting the coupling losses, as shown in Fig. 3(c). The characterization measurements are conducted using a passive optical component testing platform (detailed in Methods). At 1550 nm, the insertion losses for the four channels of the circuit are approximately 0.94 dB, 2.04 dB, 2.08 dB, and 2.48 dB, respectively. The corresponding measured crosstalks at 1550 nm are −20.09 dB, −19.21 dB, −18.02 dB, and −18.25 dB, respectively.

The micrograph of the fabricated five-mode MDM circuit is shown in Fig. 3(b), where the input and output grating coupler arrays are arranged equidistantly at 127 μm intervals, facilitating subsequent coupling and packaging with fiber arrays. Transmission spectra for the five-mode MDM circuit are obtained by launching the fundamental TE mode from ports $I_1$ to $I_5$, as shown in Fig. 3(d)-(h). At 1550 nm, the measured insertion losses for the five mode channels of the circuit are 2.96 dB, 3.62 dB, 2.43 dB, 3.94 dB, and 3.10 dB, with crosstalks of −21.73 dB, −20.72 dB, −21.91 dB, −23.43 dB, and −24.51 dB, respectively. Moreover, the transmission spectra display remarkable flatness across the entire C-band, validating its excellent potential for multi-dimensional multiplexing in on-chip optical interconnects that integrate DWDM and MDM technologies. The measurement results of the six-mode MUX for scalability validation are provided in Supplementary Note 7.

## High-speed single-$\lambda$ MDM transmission

To evaluate the capacity of the DM-based MUX for parallel data transmission across different orthogonal modes, we first conduct a high-speed intensity-modulation direct-detection (IM/DD) single-$\lambda$ MDM transmission experiment with the fabricated five-mode MDM circuit. The conceptual diagram of the interconnect setup is shown in Fig. 4(a), which consists of the E/O module, the MDM chip, and the O/E module. A single laser is used to generate continuous-wave light in this experiment. For enhanced testing stability, the chip is packaged with a pair of fiber arrays for vertical coupling (see Supplementary Note 6 for detail), with its macrograph presented in Fig. 4(b). To assess the impact of inter-mode crosstalk, we compare MDM transmission with single-port transmission. During the MDM transmission experiments, we simultaneously load optical signals into all input channels of the MDM chip, ensuring that all five modes carry decorrelated data concurrently. For each mode channel under test, it serves as the signal branch, while the remaining four form the crosstalk branch. In the contrast, the single-port configuration connects only the channel under test to the chip, while the crosstalk branch remains disconnected. As a result, data is carried exclusively by the tested mode, with no interference from other channels. Modulation and propagation losses are compensated by erbium-doped fiber amplifiers (EDFAs) for both branches. At the receiver end, the output optical signal from the mode channel corresponding to the signal branch is converted to the electrical signal by a high-speed photodiode (PD) and captured by an oscilloscope (OSC). The detailed experiment set-up is provided in Methods.

Fig. 4(d) displays the bit-error-rate (BER) performance for each channel during the transmission of 68 GBaud 8-ary pulse amplitude modulation (PAM-8) signals, as the received optical power (ROP) before the PD decreases from 4 to −1 dBm. As the received optical power drops, the performance gap between the two cases gradually narrows. This convergence is attributed to the lower signal-to-noise ratio (SNR) at reduced power levels, causing both scenarios to become noise-dominated. Experimental results show that MDM transmission incurs only a modest power penalty at the Open forward error correction (OFEC) threshold[50], with values of 0.65, 0.35, 1.0, 1.1, and 0.92 dB for MDM CH1 to CH5 respectively, demonstrating effective crosstalk suppression by our device. To further explore the capacity limits of single-$\lambda$ MDM transmission, we analyze the BER performance of PAM-8 signals under different symbol rates ranging from 48 to 128 GBaud (see Fig. 4(e)). Under the same FEC threshold, we successfully realize the transmission of 108 GBaud PAM-8 signals for all five channels, with the corresponding eye diagrams displayed in Fig. 4(f). Owing to the high channel uniformity, minimal insertion loss, and low inter-mode crosstalk of the inverse-designed five-mode MUX, we achieve the highest single-wavelength, single-mode transmission rate of 324 Gb $s^{-1}$ in on-chip MDM interconnects, amounting to an aggregate line rate of 1.62 Tb $s^{-1}\lambda^{-1}$.

**Ultra-high-capacity multi-dimensional interconnect**

To validate ultra-high-capacity on-chip optical interconnects, we employ high-order mode multiplexing combined with the spectral resources of the C-band. We utilize 88 DWDM channels on a dense 50 GHz ITU grid, covering the wavelength range from 1529.94 to 1564.68 nm. It is critical to note that integrating DWDM into MDM does not simply involve multiplying the single-$\lambda$ data rate by the number of wavelength channels. For a rigorous demonstration of multi-dimensional on-chip interconnect, it is essential to ensure simultaneous transmission of all wavelength and mode channels, which inevitably constrains the optical power per channel and introduces inter-channel crosstalk. A schematic diagram of the multi-dimensional interconnect scheme is shown in Fig. 5(a). In this experiment, we simultaneously transmit five modes, each accommodating 88 wavelength channels. Among these, three wavelength channels carry data, while the remaining 85 are filled with amplified spontaneous emission (ASE) noise to emulate a full channel loading scenario[20,51,52]. Specifically, the aggregated source module in our experiment generates the multi-$\lambda$ optical signal at the transmitter side which comprises three parts: target signal, adjacent crosstalk signal, and ASE noise channels. The target channel and its adjacent channels are generated by a tunable laser and modulated by external modulators. High-power ASE noise is generated using cascaded EDFAs and a wavelength selective switch (WSS) to flatten the ASE spectrum and carve out a slot for the laser-generated channels (see Fig. S22). The spectrum of the combined DWDM signal is shown in Fig. 5(b), where the blue background area indicates the laser-generated signal channels and the purple area corresponds to the ASE noise used to fill the C-band. During the experiment, we measure the transmission performance of all 88 wavelength channels by gradually blue-shifting the slot excavated by the WSS and the corresponding wavelengths of tunable lasers (see Methods for detailed experimental setup). We repeat this procedure for all five mode channels to comprehensively assess the performance across all 440 data channels.

Given the wavelength sensitivity of grating couplers, the loss of the packaged chip fluctuates among different channels. Therefore, the constellation probability shaping (PS) technique[53,54] is implemented to achieve flexible information rate and entropy adjustment for different mode and wavelength channels by optimizing the constellation distribution towards a Maxwell-Boltzmann distribution (see Fig. S23). In order to maximize the single-channel rate, we seamlessly switch among PAM-2, PS-PAM-4, and PS-PAM-8 formats to modulate each channel at a symbol rate of 47 GBaud with a roll-off factor of 0.05, corresponding to utilization of 98.7% of the channel spectrum. Both the achievable information rate (AIR) and net data rate (NDR) are recorded to quantify transmission performance and communication capacity (see Methods for detailed expressions for calculating date rates). Measured NDRs and entropies for all 5×88 channels are shown in Fig. 5(c), with Fig. 5(d) presenting the total AIR (7.3 to 9.1 Tb $s^{-1}$) and NDR (6.8 to 8.6 Tb $s^{-1}$) across all wavelength channels for each mode. The overall spectral efficiency reaches 8.68 bits $s^{-1}Hz^{-1}$. Ultimately, our experimental demonstration of the on-chip DWDM-MDM transmission with 440 channels achieves record-setting total AIR and NDR of up to 40.8 Tb $s^{-1}$ and 38.2 Tb $s^{-1}$, respectively, validating the capability of our inverse-designed DMs for supporting ultra-high-capacity optical interconnect.

## Discussion

Since Frellsen et al. inversely designed a dual-mode MUX on the SOI platform[46], significant efforts[9,26,35,38,39,45,47] have focused on using inverse design techniques to increase the number of modes within a compact footprint. Here we introduce a novel metric named mode density factor (MDF), defined as the device area divided by the square of the number of supported modes, for evaluating the efficiency of mode MUXs in utilizing physical space to handle multiple orthogonal modes. Table 1 provides a comprehensive literature overview of state-of-the-art fabricated mode MUXs using different methods. In recent years, the brute-force DBS method has been extensively applied to the design of DM-based mode MUXs. For instance, Liu et al. demonstrated single-port transmission of 112 Gb $s^{-1}$ discrete multi-tone (DMT) signals for each mode in a three-mode system[9]. Furthermore, Yang et al. achieved the inverse design of a four-mode MUX based on AMs, yielding an on-chip signal transmission of 1.12 Tb $s^{-1}$ using WDM-MDM technologies[26]. However, the low fabrication tolerance of AMs results in significant performance variation across devices on different chips. By merging the advantages of both approaches, we efficiently design a series of DM-based mode MUXs. Utilizing a relatively large MFS of 120 nm, our four-mode MUX achieves a remarkably low loss of 1.24 dB, which is the lowest among DM designs. Here, the insertion loss of a single MUX is approximated as half of the measured loss of the back-to-back circuit, attributed to reciprocity. Furthermore, the square hole structure in our design better complies with foundry design rule check requirements than the circular structure[9,35,38,45]. On the other hand, compared to state-of-the-art forward-designed counterparts on the SOI platform[22,23], our inverse-designed five-mode MUX not only features less stringent requirements on MFS but also reduces the overall footprint by an order of magnitude, while achieving significantly lower ILs. The EG-ADO method can extend beyond the specific devices discussed, offering potential applications in the inverse design of other sophisticated and large-scale devices, including dual-polarization mode MUXs (see Supplementary Note 8), and multi-channel wavelength

MUXs[55].

In conclusion, we have proposed a highly efficient EG-ADO method, achieving the inverse design of DM-based mode MUXs supporting up to six modes with a foundry-compatible MFS. We experimentally validate the superior performance of the four-, five-, and six-mode MUXs. Notably, the fabricated five-mode MUX supports the highest single-mode, single-wavelength transmission rate of 324 Gb $s^{-1}$ (total line rate of 1.62 Tb $s^{-1}\lambda^{-1}$) in on-chip MDM optical interconnect, and a record-setting ultra-high-capacity 38.2 Tb $s^{-1}$ multi-dimensional interconnect using 440 signal channels. Crucially, the number of parallel channels is not fundamentally limited. By employing wavelength-insensitive edge couplers[56], the estimated capacity of the multi-dimensional interconnect could reach 0.218 Pb $s^{-1}$ using 250 DWDM channels in the 1500–1600 nm waveband (see Supplementary Note 6). Furthermore, higher-order mode multiplexers can further increase the channel count with minimal additional loss and footprint (see Supplementary Note 7). Higher integration can also be achieved using integrated optical frequency combs[19,57], microring modulators[17,58], multimode routing and coupling technologies[59], enabling energy-efficient, compact, and flexible multi-dimensional on-chip interconnect across various hierarchical levels (see Supplementary Note 8). This work is expected to pave the way for next-generation on-chip optical interconnects for datacenters and hyper-scale computing systems.

## Methods

### Objective function

The objective function in inverse design is crucial as it guides the optimization process and should be specifically tailored to the intended functionality. The inverse design of a mode MUX can be regarded as a multi-objective optimization problem. The defined objective function ($L$) should include the transmission coefficients of different modes and can be expressed as follows:

$$L = 1 - \frac{1}{M}\sum_{i=1}^{M} \left\| \frac{T_i(\lambda)}{\sqrt{\Delta\lambda}} \right\|_F \tag{1}$$

Here, $M$ represents the number of modes, $T_i(\lambda)$ denotes the conversion efficiency of the fundamental TE mode from the $i$-th channel to the $TE_{i-1}$ mode in the multimode waveguide, $\Delta\lambda$ indicates the wavelength range considered for optimization, and $\|\cdot\|_F$ represents the Frobenius norm. The EG-ADO method is employed to minimize the objective function during the analog and digital optimization stages.

### Method for estimating the computational complexity

Due to variations in speed among different computational processors and simulation software, we quantify the computational complexity of the optimization process by counting the number of Maxwell solves required. In the EG-ADO method:

1. The first step involves TO using the adjoint method, where each iteration needs two Maxwell solves for forward and adjoint calculations.
2. In the third step, where DBS is used to determine the material of TBD pixels, each TBD pixel also requires two Maxwell solves for Si and $SiO_2$ cases.

Given the number of modes $M$, the number of iterations $N_A$ in TO, the total number of pixels $N_P$ in the digital pattern, and the conversion rate $C$ in the index mapping process, the total RNMS of EG-ADO method is calculated as $2M \cdot [N_A + (1-C)N_P]$ .

### Method for calculating the data rates

The AIR is determined by the normalized generalized mutual information (see Fig. S25). This metric is used to quantify the data transmission efficiency that can be achieved by optimizing the probability distribution of the signal under specific channel conditions and can be expressed as:

$$AIR = [H - (1 - NGMI) \cdot log_2(M)] \cdot R_B \tag{2}$$

where $H$ represents the signal entropy, $M$ is the modulation order, and $R_B$ is the baud rate. The NDR represents the rate at which data can be transmitted error-free. It is calculated by subtracting the FEC overhead based on the NGMI threshold[60], which can be written as:

$$NDR = [H - (1 - CR) \cdot log_2(M)] \cdot R_B \tag{3}$$

where $CR$ presents the overall code rate of the FEC coding scheme for the corresponding NGMI threshold.

**Device fabrication**

The proposed four-, five-, and six-mode MUXs are fabricated on the SOI platform, at the ShanghaiTech Quantum Device Lab of ShanghaiTech University. An electron-beam lithography (EBL) system (Elionix ELS-F125G8) is used to define the patterns and an inductively coupled plasma (ICP) etching process (HAASRODE-E200A) is applied to transfer the mask to the silicon core layer. Then the $SiO_2$ cladding layer is deposited on the silicon layer via plasma-enhanced chemical vapor deposition (Plasma Pro 100 PECVD 180). The thickness of the core layer and the cladding layer is 0.22 and 1 μm, respectively.

**Measurement set-up**

The crosstalk and insertion loss of the fabricated devices are determined through characterization measurements. The measurement setup includes a continuously tunable laser (EXFO T100S), a polarization scrambler (EPS1000), an electrically controlled fiber-chip coupling platform, and a passive optical component testing platform (EXFO CTP10). During the measurement, the fiber and grating couplers are first aligned at a fixed wavelength of 1550 nm until maximum transmission power is achieved. Subsequently, wavelength scanning is performed using the tunable laser, covering a range from 1503 nm to 1597 nm at a scanning speed of 20 nm $s^{-1}$ and an output power of 10 dBm. With the assistance of the polarization scrambler, manual adjustment of the polarization controller (PC) is unnecessary. Finally, the transmission spectra are obtained using the CTP-10.

**Single-$\lambda$ High-speed transmission experiment set-up**

The experiment set-up for single-$\lambda$ transmission is provided in Fig. S21(a). The analog electrical signal is directly loaded onto a 60-GHz thin-film lithium niobate Mach-Zehnder modulator (MZM, NOEIC MZ135-LN60) utilizing a high-bandwidth arbitrary waveform generator (AWG, Keysight M8199B) with a sampling rate of 224 GSa $s^{-1}$. A continuous-wave light, generated by an external cavity laser (Keysight N7714A) at a wavelength of 1552.5 nm and a power of 16 dBm, is fed into the MZM. The modulated signal is split into two branches by a 1×2 fiber coupler, labeled as the signal and crosstalk branches, respectively. The signal branch is amplified by an erbium-doped fiber amplifier (EDFA-I) to compensate for the optical loss introduced by the MZM. In the crosstalk branch, the optical signal is amplified by the EDFA-II, then evenly divided into four paths by a 1×4 fiber coupler. Note that the driving current for EDFA-II is set higher than that for EDFA-I to ensure that all five channels achieve a uniform input optical power of approximately 20 dBm. Each channel in the crosstalk branch incorporates a fiber delay line (DL) of varying lengths to decorrelate data across different ports. The lengths of the four fiber DLs for channel decorrelation are 0.5, 1, 2, and 5 km, respectively. At the receiver end, the output optical signal from the mode channel corresponding to the signal branch is converted to the electrical signal by a high-speed 100 GHz PD (Finisar XPDV4121R) and captured by a 256 GSa $s^{-1}$ oscilloscope (Keysight UXR0594BP) without any electrical amplifier. Detailed digital signal processing (DSP) flows for both the transmitter and receiver are provided in Fig. S21(c). To evaluate different mode channels under consistent optical signal-to-noise ratio (OSNR) conditions, manual switching of the fiber connections at the input and output ports of the packaged chip is required. For example, in the transmission test for CH1, the input fiber connector corresponding to CH1 on the packaged MDM chip is connected to the signal branch, while the input fiber connectors for CH2-CH5 are sequentially connected to the four fibers in the crosstalk branch. For subsequent measurements of CH2, the signal branch fiber is switched to CH2, while the original CH1 connector is redirected to the crosstalk branch. Corresponding adjustments are required for the fiber connections at the receiving end. Note that each switch needs recalibration of the PCs to maximize input optical power of all channels for avoiding inadvertent reductions in crosstalk.

**Multi-dimensional ultra-high-capacity interconnect experiment set-up**

The detailed experiment set-up for multi-dimensional transmission is provided in Fig. S21(b). The target signal channel generated by Laser-I and the combined light from Laser-II and Laser-III are first modulated by MZMs with a 3-dB bandwidth over 20 GHz and a $V_\pi$ of 5.5 V (Sumicem T.MXH 1.5-20PD-ADC-LV). These MZMs are driven by analog signals generated by a 120 GSa $s^{-1}$ AWG (Keysight M8194A) at a voltage swing of 200 mV, which are then amplified via electrical amplifiers. The ASE signal, initially at a low power of 19 dBm from an ASE/EDFA (Amonics AEDFA-23-B-FA), is spectrally shaped by a WSS (Finisar Waveshaper 4000S) and then amplified to 26 dBm by a gain-flattening ASE/EDFA (OVLINK EDFA-C-BA-GF-26-PM-B). This WSS is placed between two EDFAs to prevent damage from excessive power levels. The generated ASE noise is first merged with the adjacent signal channels via a 1×2 polarization-maintaining fiber coupler (PM-FC), and this combined light is then coupled with the target signal channel through a 2×2 PM-FC. The outputs from this coupler are fed into the MDM chip as the signal and crosstalk branches, respectively. EDFA-I is used to ensure that the power level in each channel of the crosstalk branch matches that of the signal branch (~18 dBm). The setup of fiber DLs and PCs is identical to that used in the single-$\lambda$ experiment. At the receiver end, the output from the MDM chip is first passed through an OBPF (EXFO XTM-50-SCL-U) to extract the target signal channel. This signal is then amplified to approximately 5 dBm by EDFA-II (OVLINK EYDFA-C-HP-BA-30-PM-B) before being received by a high-speed PD and OSC. DSP for both the transmitter and receiver sides is conducted offline. During the generation and changing of the WDM signals, an optical spectrum analyzer (OSA, Yokogawa AQ6370D) is used to monitor and align the 50-GHz-grid DWDM channels. Each change in the WDM channel necessitates reloading the WSS configuration file written in MATLAB, adjusting the laser frequencies, and tuning the OBPF wavelength knob. Manual switching of the input fibers is performed for testing different mode channels.

## Data Availability

The source data generated in this study have been deposited in the Figshare database under accession code https://doi.org/10.6084/m9.figshare.27074551.

## Code Availability

The open-source code of the inverse design algorithm is available at https://github.com/BryvinSal/EGADO. Cite the current article for using the code.

## Acknowledgements

We thank J. Hu and Y. Li for assistance with the image preparation. This work is partly supported by the National Key Research and Development Program of China (No. 2023YFB2905700), and the National Natural Science Foundation of China (Nos. 62235005, 62171137 and 61925104).

## Author contributions

A.S. conceived the basic idea. J. Zhang and F.H. contributed to the basic framework and feasible technical route of the project. A.S. and S.X. designed specific research content according to the route. A.S. designed and proposed the edge-guided analog-and-digital optimization method with the assistance of X.D and Y.Y. R.S. and A.S. completed the device fabrication with the assistance of B.H. A.S., A.Y. conducted the single-wavelength transmission experiments with the assistance of B.D. A.S., S.X., and A.Y. conducted the multi-dimensional transmission experiments with the assistance of O.H. and J. Zhao. J. Zhang and N.C. guided the adjustment of experiments. Z.L., J.S., Y. Zhou, C.S., and Y. Zhao participated in preparing the manuscript and contributed to the discussions. A.S. wrote the manuscript and revised it based on comments from all authors. The project was performed under the supervision of J. Zhang, N.C., W.C., and H.C.

## Competing interests

The authors declare no competing interests.

# Tables

Table 1. Performance comparison of experimentally demonstrated mode MUXs on SOI

| Year/ Reference | Mode Count | IL (dB) | MFS (nm) | CT (dB) | Area ($\mu m^2$) | MDF ($\mu m^2$) | Algorithm | Method | Data Rate |
|---|---|---|---|---|---|---|---|---|---|
| 2017[25] | 3 | 0.2 | 120 | -20 | 3.6×310 | 124 | ADC | | N.A. |
| 2021[24] | 4 | 1.8 | 125 | -15.5 | 4.2×49[a] | 12.86 | TWC | Forward-designed | N.A. |
| 2023[23] | 5 | 3.4 | 100 | -15 | 12.3×60[a] | 29.52 | ADC | | N.A. |
| 2023[22] | 5 | 3.8 | 100 | -20 | 18.3×32.6[a] | 23.86 | ADC | | N.A. |
| 2021[39] | 2 | 0.9 | 90 | -15 | 2.4×3 | 1.8 | DBS | | 25 Gb $s^{-1}$×2[b] |
| 2018[40] | 2 | 1.0 | 90 | -24 | 2.4×3 | 1.8 | DBS | | N.A. |
| 2016[46] | 2 | 1.2 | 32 | -12 | 2.6×4.22 | 2.74 | TO | | N.A. |
| 2020[47] | 2 | 1 | 90 | -15.6 | 2.55×3.55 | 2.26 | TO | | N.A. |
| 2019[9] | 3 | 1.0 | 120 | -20 | 3.4×3.9 | 1.47 | DBS | | 112 Gb $s^{-1}$×3[b] |
| 2022[35] | 4 | 1.48 | 90 | -13 | 4.8×4.8 | 1.44 | DBS | Inverse-designed | N.A. |
| 2020[38] | 4 | 1.5 | 130 | -19 | 5.4×6 | 2.02 | DBS | | N.A. |
| 2021[45] | 4 | 1.4 | 120 | -15 | 6×6.8 | 2.55 | DBS | | N.A. |
| 2022[26] | 4 | 0.5 to 3[c] | 80 | -13 to -18[c] | 6.5×6.5 | 2.64 | TO | | 1.12 Tb $s^{-1}$ |
| This work | 4 | 1.24 | 120 | -18 | 6×6 | 2.25 | EG-ADO | | N.A. |
| This work | 5 | 1.97 | 120 | -20 | 6×10 | 2.4 | EG-ADO | | 38.2 Tb $s^{-1}$ |

IL, insertion loss; MFS, minimum feature size; CT, crosstalk; MDF, mode density factor; ADC, adiabatic directional coupler; TWC, triple-waveguide coupler; DBS, direct binary search; TO, topology optimization; N.A., not available. [a]Actual footprint obtained from the SEM image. [b]Transmission experiments based on single-port operation. [c]Not given directly in the literature but can be observed from the figure.

# Figures

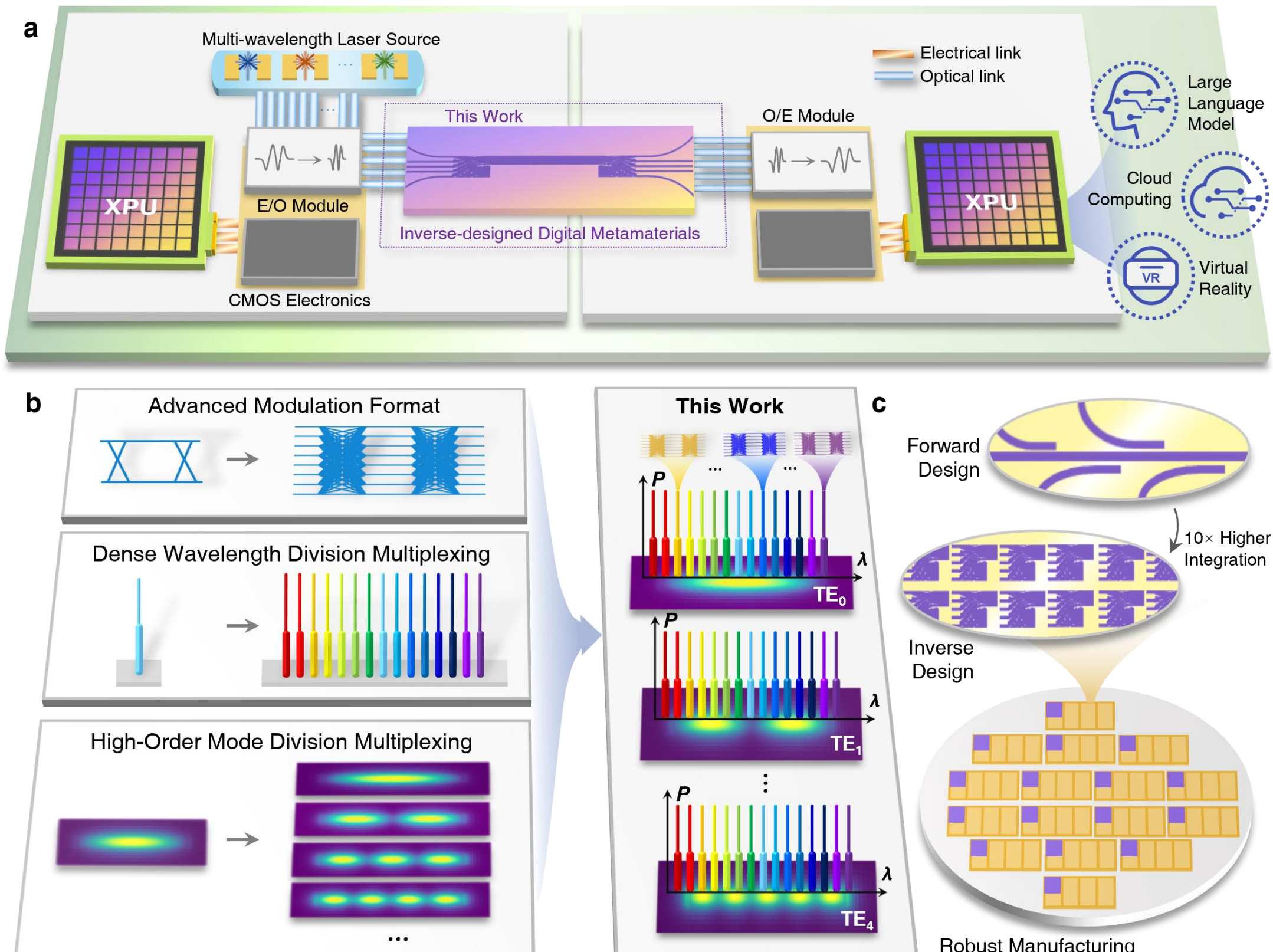

**Fig. 1 On-chip multi-dimensional interconnect architecture. a** Artistic representation for an on-chip link between two XPUs utilizing inverse-designed digital metamaterials within future optical compute interconnect systems. Here, the term "XPU" serves as a device abstraction including various computational architectures, including CPUs, GPUs, FPGAs, and other accelerators. These broadband, inverse-designed digital metamaterials facilitate the multiplexing of a substantial number of data channels across both wavelength and mode dimensions. Additionally, the system incorporates E/O module for converting electrical signals to optical signals, and O/E module for the reverse conversion. Signal processing can be supported by co-packaged CMOS electronics. **b** Strategies on enhancing interconnect capacity across three dimensions: symbol, wavelength, and mode. This study showcases the integration of MDM and DWDM, enabling nearly a hundred wavelength channels per orthogonal mode. Each channel efficiently supports high-order pulse amplitude modulation (PAM) signals, significantly expanding the data capacity per waveguide. **c** Foundry-compatible inverse-designed digital metamaterials. Our design features a minimum feature size of 120 nm, suitable for robust and large-scale production by commercial foundries. Compared to conventionally designed high-order mode MUX, the inverse-designed MUX significantly reduces the footprint by an order of magnitude.

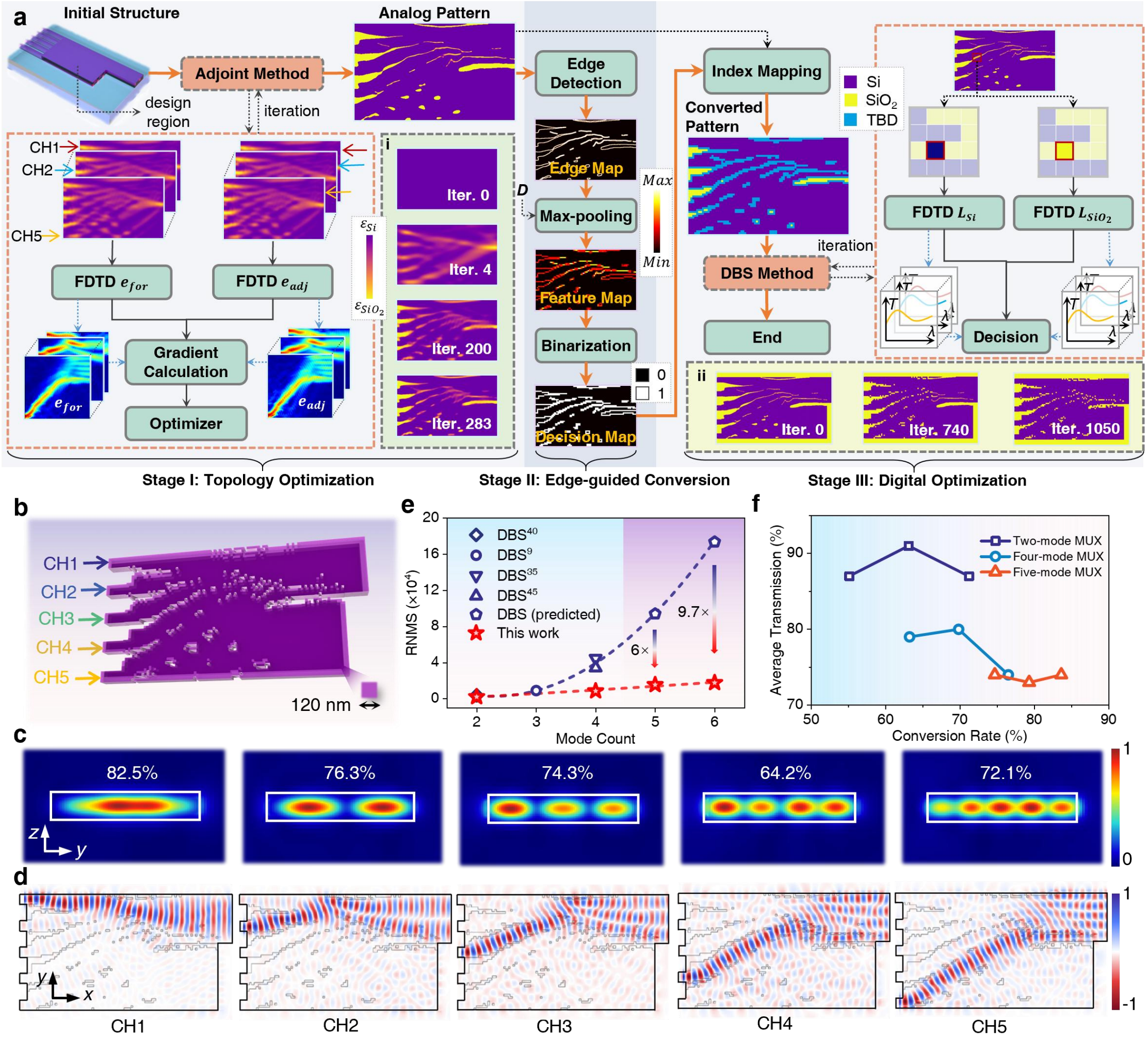

**Fig. 2 The EG-ADO method and the DM-based five-mode MUX. a** The workflow of the EG-ADO method. During each iteration in the first stage, the forward electric fields ($\mathbf{e_{for}}$) and adjoint fields ($\mathbf{e_{adj}}$) for all modes are used to obtained the pixel gradient. During the second stage, a dilation parameter (*D*) is introduced to regulate the direct conversion rate from analog to digital metamaterials. During the final stage, the TBD pixels are iteratively traversed column by column. The objective functions for Si ($L_{Si}$) and $SiO_2$ ($L_{SiO_2}$) configurations are compared for determining the final pixel material. For any TBD pixels not yet determined, their state remains equivalent to that of the analog pattern. Insets (i) and (ii) depict the evolution of the index pattern for the five-mode MUX during topology and digital optimization stages, respectively. **b** Final structure of the DM-based five-mode MUX. The pixel side length is 120 nm. **c** Simulated mode profiles at the output multi-mode waveguide for five channels at the wavelength of 1550 nm. The power conversion efficiency is provided along with each 2D profile. **d** Simulated light propagation process for five channels at the wavelength of 1550 nm. **e** Computational complexity comparison of state-of-the-art inverse-designed DM-based mode MUXs on the SOI platform. Dashed lines represent the fitted curves. The purple-shaded area denotes the regime that has not been reported in the literature due to the high computational complexity and time consumption of conventional DBS methods. **f** Average transmission of all modes for two-, four-, and five-mode MUX designed by EG-ADO method under different conversion rate at the wavelength of 1550 nm. For each MUX, changes in the conversion rate are realized by adjusting the dilation parameter *D*, with values set at *D* = 0, 2, and 4. A higher *D* value corresponds to a lower conversion rate due to an increased count of pixels designated as TBD during the index mapping process.

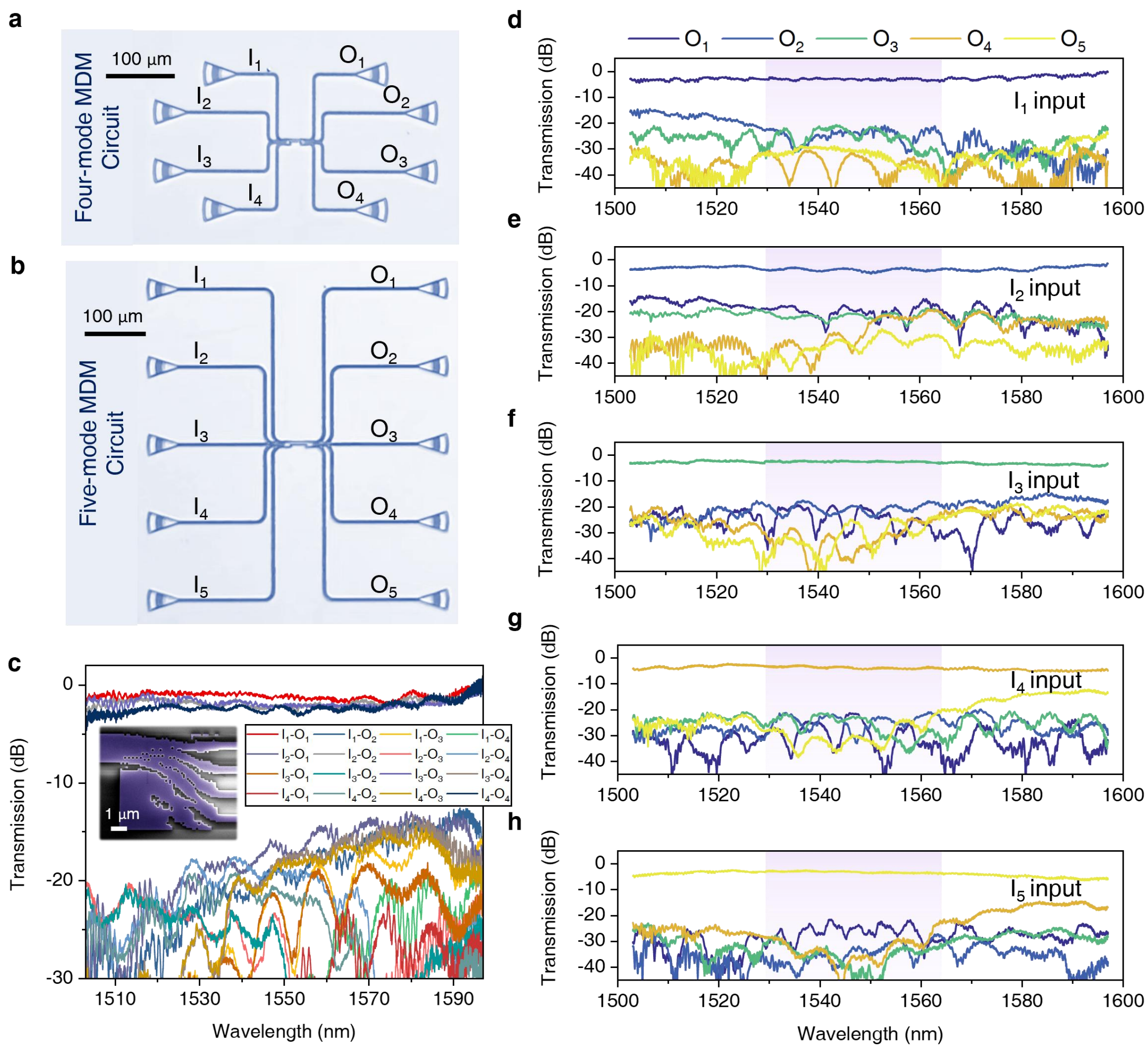

**Fig. 3 Measured optical performance of the fabricated four-mode and five-mode MUXs designed with the EG-ADO method. a-b** Microscope images of the back-to-back **(a)** four-mode and **(b)** five-mode MDM circuits. **c** Measured transmission spectra of the four-mode circuit. The inset depicts the false-color scanning electron microscopy (SEM) image of the four-mode MUX. **d-h** Measured transmission spectra of the five-mode circuit when the light is incident from $I_1$ to $I_5$. The shaded areas represent the C band range (1530-1565 nm).

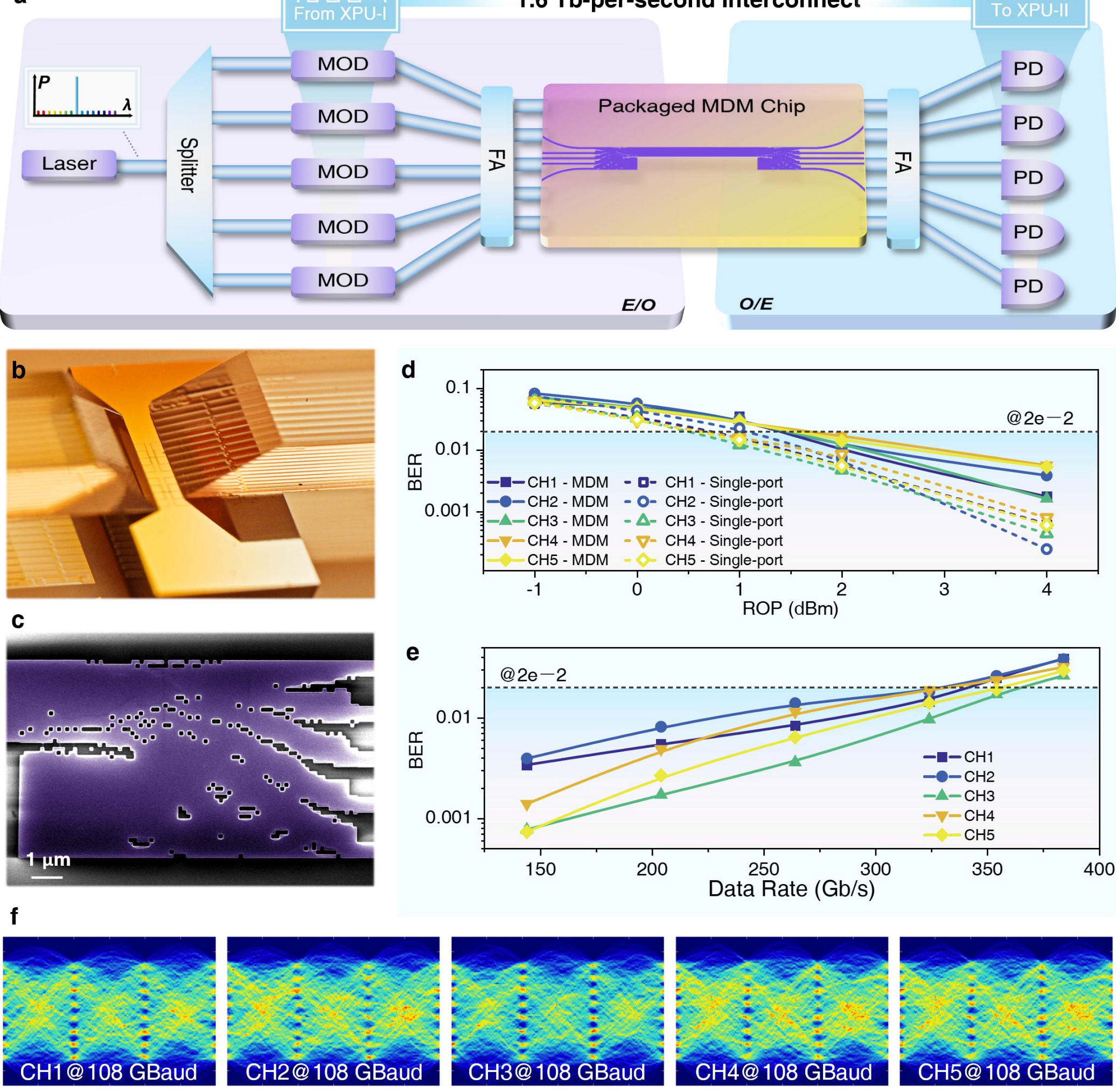

**Fig. 4 High-speed single-$\lambda$ MDM on-chip interconnect. a** Schematic of the single-$\lambda$ MDM transmission based on the packaged five-mode chip. Input continuous-wave light is divided by a splitter into five paths, each modulated by a modulator (MOD) and then coupled to a packaged chip via a fiber array (FA). Each path corresponds to an orthogonal mode in the MDM chip. At the receiver end, the mixed signal is demultiplexed by a mode deMUX and then coupled into photodiodes (PD) through another fiber array, where different mode channels are received and demodulated individually. **b** Macroscope image of the packaged MDM chip with a pair of fiber arrays. **c** False-color SEM image of the five-mode MUX. It confirms the successful transfer of the expected digital pattern onto the chip. Etched holes with well-defined square shapes and point-like contacts along diagonals indicate a high-quality fabrication result that accurately replicates the target design. This fabrication robustness stems from our digitally designed pattern and its relatively large MFS. **d** BER measurements for the MDM and single-port transmission for all five mode channels under different ROPs. An indicative FEC threshold is given at 0.02, corresponding to a pre-FEC BER for the OFEC threshold. BER, bit error rate; ROP, received optical power. **e** BER measurements for MDM transmission for all five channels under different data rates. At the data rate of 324 Gb $s^{-1}$, all five channels exhibit BER values below the OFEC threshold. **f** Corresponding eye diagrams for the 108 GBaud PAM-8 signals at different channels.

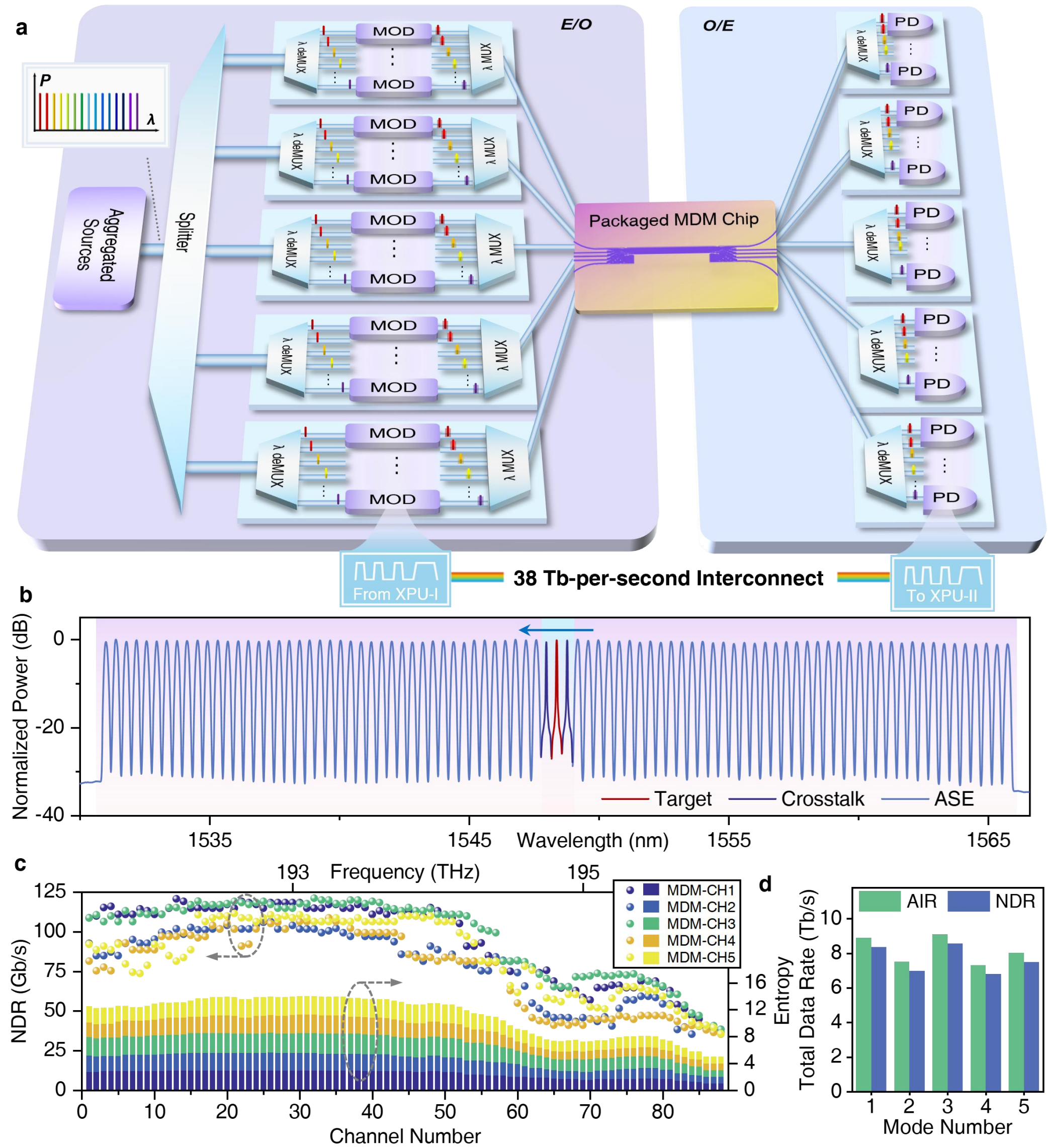

**Fig. 5 Ultra-high-capacity on-chip multi-dimensional interconnect. a** Schematic of the multi-dimensional on-chip interconnect scheme. The spectrally flattened multi-wavelength signal, generated from aggregated sources, is distributed into five branches corresponding to different orthogonal modes on the chip. Each branch utilizes a wavelength demultiplexer ($\lambda$ deMUX) to segregate different wavelength channels for independent modulation. Subsequently, a wavelength multiplexer ($\lambda$ MUX) reassembles the WDM signal. Fiber arrays (not depicted for clarity) link the packaged chip with E/O and O/E modules. At the receiving end, the WDM signals are extracted by $\lambda$ deMUXs and converted into electrical signals by photodiodes. In this experiment, 88 wavelength channels across 5 waveguide modes are employed, facilitating the propagation of a total of 440 parallel signal lanes on-chip. The entropy of each lane is flexibly adjusted based on channel conditions. **b** Normalized optical spectrum of the combined DWDM signal at the transmitter side. Here, the 44-th channel is modulated with target signal by a Mach-Zehnder modulator (MZM), while its adjacent channels are driven by another MZM. The remaining wavelength channels in the C band are filled with the ASE noise to ensure full spectral utilization. **c** Measured net data rates (left axis) and entropies (right axis) across 88 wavelength channels for MDM CH1 to CH5. The frequency range of the DWDM signal corresponds to 191.6 to 195.95 THz. **d** Total AIR and NDR for five mode channels. AIR, achievable information rate; NDR, net data rate.